\documentclass[twocolumn]{article}
\pdfoutput=1
\usepackage{graphicx,amsmath,amssymb,textcomp,fullpage,url}
\usepackage[version=4]{mhchem}
\usepackage{setspace}
\usepackage{booktabs}

\graphicspath{ {./Figures/} } 

\usepackage{color}
\usepackage[normalem]{ulem}

\begin{document} 

\title{Structural and dynamic disorder, not ionic trapping, controls charge transport in highly doped conducting polymers}

\author{Ian E. Jacobs,$^1$ Gabriele D'Avino,$^2$ Vincent Lemaur,$^3$ Yue Lin,$^1$ Yuxuan Huang,$^1$\\  Chen Chen,$^1$ Thomas Harrelson,$^4$ William Wood,$^1$ Leszek J. Spalek,$^1$ Tarig Mustafa,$^{1,5}$\\ Christopher A. O'Keefe,$^5$ Xinglong Ren,$^1$ Dimitrios Simatos,$^{1,4}$ Dion Tjhe,$^1$ Martin Statz,$^1$\\ Joseph Strzalka,$^6$ Jin-Kyun Lee,$^7$ Iain McCulloch,$^{8,9}$ Simone Fratini,$^2$ David Beljonne,$^3$\\ Henning Sirringhaus$^1$}

\twocolumn[{
\maketitle

$^1$ Optoelectronics Group, Cavendish Laboratory, University of Cambridge, J J Thomson Avenue, Cambridge CB3 0HE, UK
$^2$ Grenoble Alpes University, CNRS, Grenoble INP, Institut Néel, 25 rue des Martyrs, 38042 Grenoble, France
$^3$ Laboratory for Chemistry of Novel Materials, University of Mons, Mons, B‐7000 Belgium
$^4$ Molecular Foundry, Lawrence Berkeley National Laboratory, One Cyclotron Road
Building 67, Berkeley, CA 94720, USA
$^5$ Department of Chemistry, University of Cambridge, Lensfield Road, Cambridge, CB2 1EW, UK
$^6$ X‐Ray Science Division, Argonne National Laboratory, Lemont, IL, 60439 USA
$^7$ Department of Polymer Science \& Engineering, Inha University, Incheon, 402‐751 South Korea
$^{8}$ KAUST Solar Center, Physical Sciences and Engineering Division (PSE), Materials Science and Engineering Program (MSE), King Abdullah University of Science and Technology (KAUST), Thuwal, Kingdom of Saudi Arabia
$^{9}$ Department of Chemistry, University of Oxford, Oxford, UK

\begin{abstract}
Doped organic semiconductors are critical to emerging device applications, including thermoelectrics, bioelectronics, and neuromorphic computing devices. It is commonly assumed that low conductivities in these materials result primarily from charge trapping by the Coulomb potentials of the dopant counter-ions. Here, we present a combined experimental and theoretical study rebutting this belief. Using a newly developed doping technique, we find the conductivity of several classes of high-mobility conjugated polymers to be strongly correlated with paracrystalline disorder but poorly correlated with ionic size, suggesting that Coulomb traps do not limit transport. A general model for interacting electrons in highly doped polymers is proposed and carefully parameterized against atomistic calculations, enabling the calculation of electrical conductivity within the framework of transient localisation theory. Theoretical calculations are in excellent agreement with experimental data, providing insights into the disordered-limited nature of charge transport and suggesting new strategies to further improve conductivities.

\end{abstract}

\par}]

\clearpage
The Nobel Prize winning discovery of high, metallic electrical conductivities in polyacetylene doped by exposure to oxidising agents\cite{Chiang:1977} initiated a strong research interest in understanding the charge transport physics of conducting polymers. This interest has recently been re-invigorated by new materials systems and emerging applications in sensing, bioelectronics,\cite{Someya:2016} and thermoelectrics.\cite{Russ:2016,Kroon:2017} Compared to inorganic metals, conducting polymers exhibit a number of unique characteristics, which make the description of their transport physics complex.\cite{Baranovskii:2014,Fratini:2020} These include a highly anisotropic electronic structure with strong covalent interactions along the polymer chain and weaker van der Waals interactions between chains; strong electron-phonon interactions reflecting the soft molecular nature and resulting in polaron formation, the presence of structural and static energetic disorder associated with torsional defects or chain ends, variations in $\pi$-$\pi$ stacking distances or generally spatial variations in chain conformation, and the influence of dynamic fluctuations of the electronic couplings between molecular units due to strong thermal lattice fluctuations and molecular vibrations. Furthermore, the doping concentrations to achieve the highest conductivities tend to be $>10^{20}$ cm$^{-3}$ and approach the density of molecular repeat units.\cite{Jacobs:2017a,Zhao:2020}  At such high densities the dopant counter-ions modify the polymer microstructure and may cause additional structural disorder. Additionally, due to the low dielectric constant characteristic of these materials, it is important to consider the strong, attractive Coulomb forces between the electronic charge carriers and the dopant ions as well as the repulsive Coulomb interactions between the carriers. 

Much progress has been made in recent years in understanding the charge transport physics of semiconducting organic systems, studied most commonly with field-effect gating at much lower carrier concentrations and in the absence of dopant ions.\cite{Noriega:2013, Fratini:2020} There is an emerging consensus that the framework of transient localisation (TL)\cite{Fratini16,Troisi:2006} provides the most appropriate description of charge transport in molecular crystal field-effect transistors (FETs) with high charge carrier mobilities of 1-20 cm$^{2}/$Vs. On timescales faster than the structural lattice dynamics TL considers charge carriers to be localized by the combined effects of static disorder and the dynamically generated configuration of site energies and transfer integrals. However, this energetically disordered landscape evolves with the lattice dynamics, and on longer timescales charge carriers are able to undergo a diffusive motion and effectively ``surf on the waves of molecular lattice distortions". The TL framework provides an explanation of many of the characteristics transport signatures of molecular crystals and is in good quantitative agreement with their experimentally observed mobility values.\cite{Fratini:2020}

\begin{figure*}[ht!]
\begin{center}
\includegraphics[width=\textwidth]{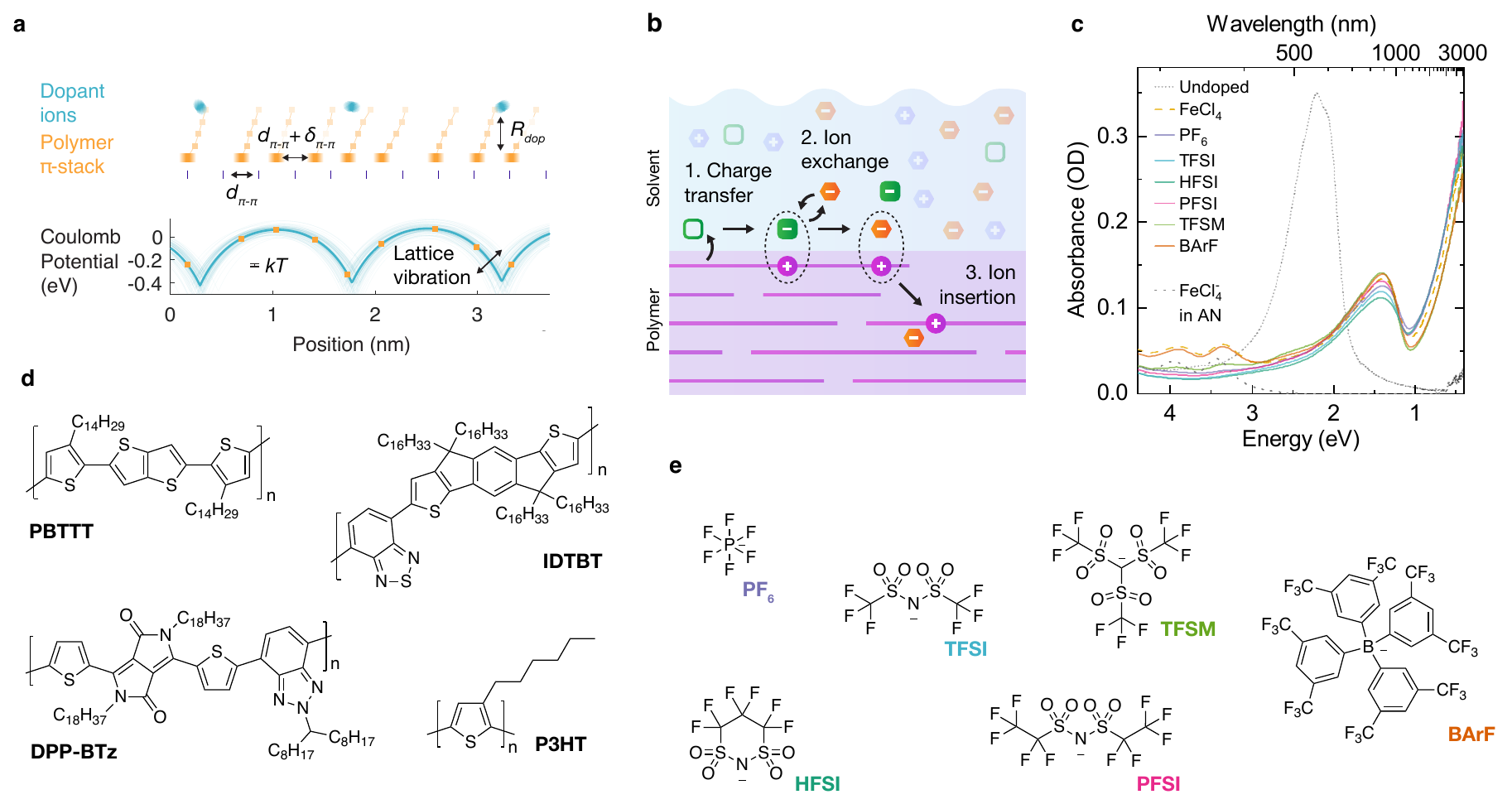}
\caption{\textbf{Ion exchange doping}. a) Schematic of a doped polymer aggregate; blue circles represent dopant ions, orange squares represent monomers; blurring represents thermal motion. Here $d_{\pi-\pi}$ is the $\pi$-stacking distance and $\delta_{\pi-\pi}$ consists of both the static disorder in stacking distances (paracrystalline disorder) and the dynamic disorder (thermal motion); $R_{dop}$ is the distance between the dopant ion and the polymer backbone plane. The lower plot illustrates the potential energy surface along the $\pi$-stacking direction resulting from the dopant ion Coulombic interaction ($R_{dop}$ = 5 \AA, $N_d$ = 25 mol\%; hole-hole interactions are not included). Thin lines illustrate the effect of vibrations on the Coulomb potential ($\delta_{rms} = 0.5$\AA); thicker line shows the equilibrium potential.  b) Schematic of the ion exchange doping process. c) UV-vis-NIR spectra of PBTTT doped with electrolytes consisting of TFSI anions with different anions (100 mM, AN), mixed with FeCl$_3$ (1 mM). Doping solution exposure time is 5 minutes, corresponding to saturation doping level d,e) Structures of polymers (d) and dopant ions (e) studied here.}
\label{Fig1}
\end{center}
\end{figure*}

An important unresolved question is whether the TL framework accurately describes transport in highly doped systems, e.g. those relevant to thermoelectric applications. The electronic structure and the ensuing transport physics of highly doped conjugated polymers is much more complex and richer than that of pristine and highly-crystalline molecular semiconductors, and its rationalization calls for several extensions. The high carrier density brings these systems into the realm of the complex many-body physics of Coulobically-interacting particles, whose effects are expected to be strongly enhanced in low-dimensional organic materials characterized by weak dielectric screening due to the low dielectric constant of the host lattice. Coulomb forces concern both the interactions among carriers on polymer chains, as well as between carriers and dopant ions. The role of carriers’ interactions in a fluctuating energy landscape has been studied very recently, with results for different materials showing that many-body phenomena effectively contribute to the energetic disorder causing the transient localization of charge carriers.\cite{Fratini:2021} One of the poorly understood questions concerns the role of Coulombic traps that the dopant ions may create in the density of states\cite{Arkhipov:2005} and to which extent these are limiting transport at high carrier densities. Although the depth of these integer charge transfer complex (ICTC) states decreases as the potential wells between adjacent ions overlap, even at high doping levels the depth of these wells remains many times greater than the thermal energy at room temperature, $kT$ (Figure \ref{Fig1}a). On the other hand, as the localization length of the charge carriers becomes comparable to the separation between dopant ions, ICTC states should delocalize into an impurity band. The Mott criterion, $N_d^{-1/3} R_{dop} \approx 0.2$, provides a rough estimate of when this delocalization might occur.\cite{Mott:1968} For an ion-polaron distance $R_{dop}$ = 4 \AA, this suggests that above a doping level $N_d$ = 10$^{20}$ cm$^{-3}$, corresponding to a molar doping level of about 10\%, dopant-polymer ICTCs may not behave as trap states, despite their large binding energy. However, this picture is very simplistic; additional factors must also be considered, particularly the ever-present role of static and dynamic disorder, illustrated in Figure \ref{Fig1}a, which strongly affects the energy of polymer sites and modulates the transfer integral between adjacent chains.\cite{Noriega:2013,Fratini:2020}

Here we report a systematic study of the influence of Coulomb traps on charge transport by varying in a controlled manner the dopant ion size and shape and the distance between the dopant ions and the charge carriers by using a recently developed ion-exchange doping technique.\cite{Yamashita:2019,Jacobs:2021a,Thomas:2020,Murrey:2021} We selected a series of four conducting polymers with different degrees of crystallinity, which allows us to investigate, in particular, the role of paracrystallinity, which has been shown to be a key factor governing the carrier mobility in FETs at low carrier concentrations and in the absence of counter-ions \cite{Noriega:2013}. However, in conducting systems paracrystalline disorder and Coulomb interactions are not independent, as the dopant ions can be the source of significant structural distortions and microstructural changes when they are incorporated into the polymer film. To explain our experimental observations we build a microscopic model of charge transport in conducting polymers within the transient localization framework, and show that theoretical predictions of the model are in excellent quantitative agreement with the experimental observations, including the values of the observed electrical conductivity and the dependence of conductivity on paracrystallinity. Our work suggests that the TL framework is applicable to describing the transport physics of state-of-the-art conducting polymers and provides new insight into the factors that limit their conductivities.

\section*{Results}

Systematic studies as a function of dopant size and shape have recently been reported for vacuum-sublimed molecular systems, in which the dopant can be co-evaporated with the host molecule. Schwarze et al.\cite{Schwarze:2019} investigated a range of doped molecular semiconductors and showed that at low doping concentrations the activation energy of the conductivity is determined by the Coulomb binding energy of the integer charge transfer complex (ICTC) between the host and dopant ions. At doping concentrations $>$10\% the activation energy is reduced and determined by the energetic disorder among the ICTC states. In polymers doped by the conventional charge transfer method such systematic studies of the role of the dopant ion are more difficult, as for a specific polymer there is usually only a very limited number of dopant molecules that provide efficient doping, and the sites into which dopant ions are incorporated cannot be controlled and remain usually undetermined. An elegant approach has been to tether the dopant to the alkyl side chains of the polymer,\cite{Mai:2014,Tang:2016} but such self-doped polymers typically do not exhibit very high conductivities and are only of limited use for understanding the factors that limit the conductivity of state-of-the art conducting polymers. 

To overcome these limitations we make use here of a recently reported method for doping based on ion exchange that opens new opportunities for performing systematic studies of dopant size and shape.\cite{Yamashita:2019,Jacobs:2021a} In conventional p-type charge transfer doping the dopant molecule needs to perform two functions: it induces the initial electron transfer from the host molecule, then is incorporated into the film as a radical anion. This limits the choice of suitable dopants. In contrast, in ion exchange doping a doping solution is used which contains both a molecular dopant and an electrolyte (Figure \ref{Fig1}b). After charge transfer onto the molecular dopant, the reduced dopant is exchanged with the negative ion of the electrolyte, which is then incorporated into the film as a stable, closed shell counterion. This provides more stable doping but also allows selecting the counterion systematically from a wide range of stable ions with different size and shapes.  Very recent work by Thomas \textit{et al.} applied the ion exchange method to the polythiophene-based, semicrystalline polymer, poly[2,5-bis(3-tetradecylthiophen-2-yl)thieno[3,2-b]thiophene] (PBTTT) and found little variation in conductivity with ion size.\cite{Thomas:2020} However, the conductivities reported for ion-exchanged PBTTT have been relatively low to date (320 S/cm in Ref. \cite{Thomas:2020}). It is important to understand the factors which limit conductivity in order to identify routes to higher performance and also to test the generality of such observations across different polymer systems.  

For our study we selected four polymers according to their microstructure ranging from semicrystalline to nearly amorphous (Figure \ref{Fig1}D). Our implementation of ion exchange doping, described previously,\cite{Jacobs:2021a} is a simple sequential doping process\cite{Jacobs:2016} applied to polymer films deposited in their undoped form. The doping solution consists of FeCl$_3$ plus a large molar excess of electrolyte in anhydrous acetonitrile. As demonstrated previously, under anhydrous conditions FeCl$_3$ is an extremely powerful oxidizing agent, stronger than any reported organic molecular dopant. The excess electrolyte (100 mM electrolyte / 1 mM FeCl$_3$) provides the entropic driving force for ion exchange.

UV-vis-NIR spectra of PBTTT doped with blends of FeCl$_3$ and various electrolytes (Figure \ref{Fig1}c) show nearly complete bleaching of the polymer $\pi-\pi^*$ band, indicating a uniformly high doping level for all ions. Results for the other three polymers studied, P3HT, DPP-BTz, and IDTBT, show similar behavior (Figure S1). Doping conditions (100/1 mM electrolyte:FeCl$_3$, AN, 5 min exposure) were chosen to allow time for diffusion of larger ions, while limiting material degradation, which we previously identified as a concern.\cite{Jacobs:2021a} Ion exchange efficiency, estimated from the residual FeCl$_4^-$ concentration extracted from fits to the UV spectra (Figure S2) shows systematic variation with respect to ionic volume and polymer crystallinity. For small and medium ions such TFSI, and HFSI, exchange efficiency is universally high, consistent with our previous characterization of PBTTT:TFSI which revealed exchange efficiencies exceeding 99\%.\cite{Jacobs:2021a} However, in the most crystalline material studied here, PBTTT, exchange efficiency drops to almost zero for the largest ion, BArF. The strong ionic size dependence of the ion exchange efficiency is consistent with our previous assertion of a strong crystalline strain contribution to the ionic selectivity.\cite{Jacobs:2021a} Further discussion of the exchange efficiency is given in Supporting Information Section 1.

\subsection*{Characterizing carrier density}

Because of the expected changes in electronic structure as doping level varies, characterization of the doping level in our samples is critical. Qualitatively, the complete bleaching of the polymer $\pi-\pi^*$ bands indicate our doping process is capable of reaching very high carrier concentrations. Because the ionization efficiency in ion exchange doped polymers is by definition 100\%, a simple method of measuring the carrier density is simply to count the number of dopant ions in the film. Unfortunately, the extremely wide optical gap of the closed-shell ions used in ion exchange doping prevents direct measurement via UV-vis spectroscopy. The absorption of TFSI, for instance, is below 200 nm, and therefore difficult to quantitatively separate from the substrate absorption edge and high-lying polymer transitions. We can, however, estimate the doping level in the absence of ion exchange, by fitting the FeCl$_4^-$ absorption in films doped with FeCl$_3$ only (yellow dashed line in Figure \ref{Fig1}c). Because ion exchange is driven by the entropy provided by the large excess of electrolyte ions, ion exchange should always make the doping reaction slightly more exothermic. Therefore, the doping levels estimated by fitting the FeCl$_4^-$ absorption without ion exchange (Table S2) is generally a slight underestimate of the value obtained with ion exchange, consistent with the findings of Yamashita et al.\cite{Yamashita:2019}

Unfortunately, a more direct measurement of the doping level in ion exchange doped materials is less straightforward. This difficulty arises from several factors and is not specific to the PBTTT:TFSI system. TFSI has many vibrational modes but these overlap strongly with the polymer IRAV modes, which again complicates quantitative measurements.\cite{Rey:1998} Spectroscopic quantification of the polymer polaron bands is also difficult in this regime due to changes in polaron delocalization at high doping levels, which in turn changes the shape and intensity of the polaron IR bands.\cite{Ghosh:2020} Electron spin resonance (ESR) can be used to measure the unpaired spin density in the material, however, at high doping levels we often observe a Pauli-type magnetic suceptibility, in which the spin density corresponds only to the subset of spins in close vicinity to the Fermi level.\cite{Kang:2016, Yamashita:2019, Tanaka:2020} The Hall effect also sometimes becomes measurable in polymers at high carrier densities.\cite{Wang:2012,Kang:2016} However, as discussed below, due to partial screening by hopping carriers the Hall coefficient typically overestimates the true carrier density.\cite{Yi:2016} These difficulties forced us to explore alternative methods of carrier density quantification. Here, we identify two methods: X-ray photoemission spectroscopy (XPS) and quantitative nuclear magnetic resonance (QNMR), both of which enable direct quantification of the TFSI ion density in PBTTT.

\begin{figure*}[ht!]
\begin{center}
\includegraphics{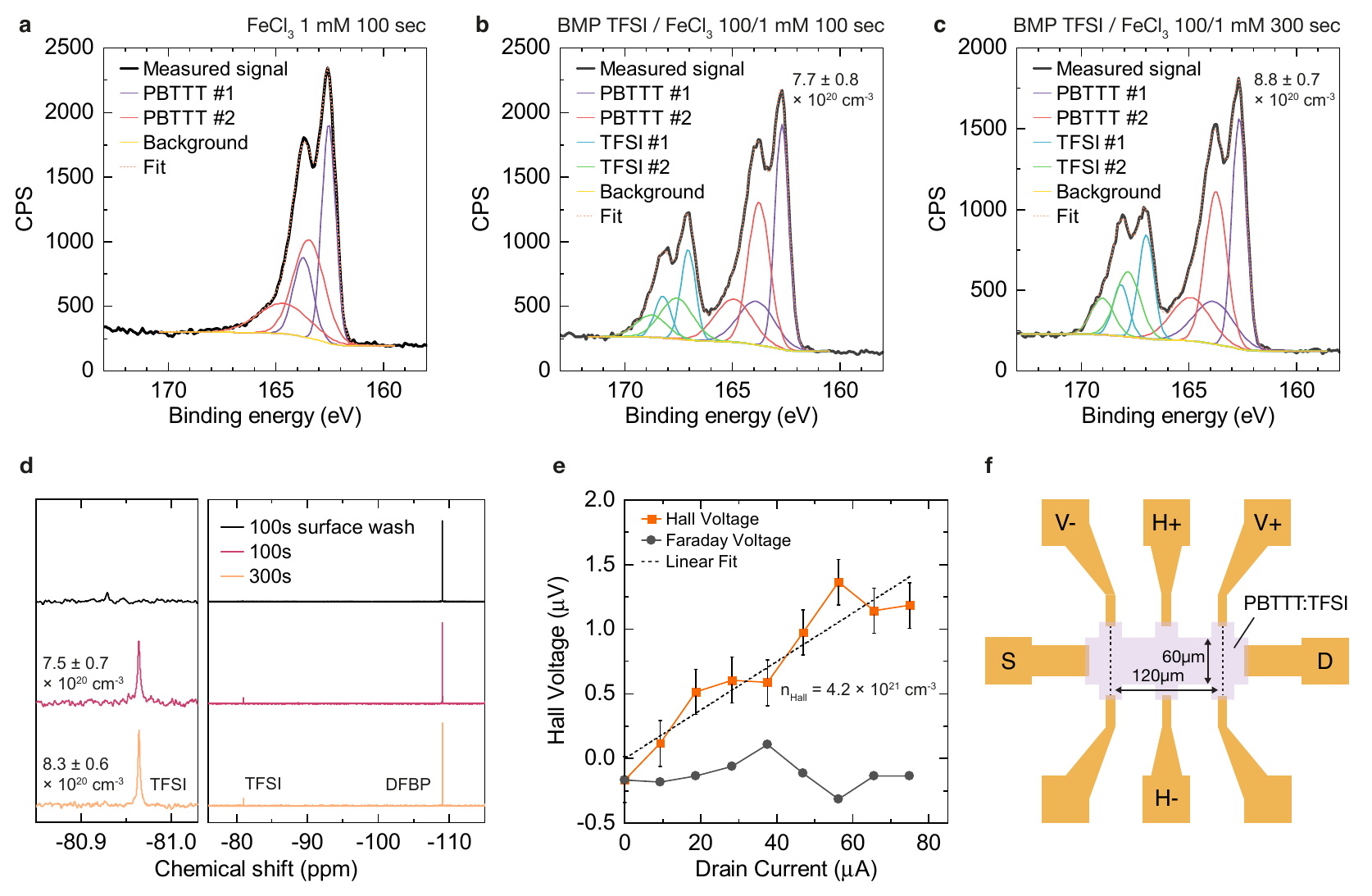}
\caption{\textbf{Carrier density measurement in PBTTT:TFSI}. Sulfur 2p XPS spectra of doped PBTTT:TFSI films: a) 1 mM FeCl$_3$, 100 sec (without ion exchange); b) 100/1 mM BMP TFSI / FeCl$_3$, 100 sec; c) 100/1 mM BMP TFSI / FeCl$_3$, 100 sec. d) $^{19}$F QNMR spectra of IEx doped PBTTT:TFSI films (100 or 300 sec exposure time, 100/1 mM BMP TFSI / FeCl$_3$ in AN) after either washing with CD$_3$CN, or dedoping wtih TEA (10\% v/v in CD$_3$CN, 5 min). Left plot shows detail of TFSI peak. e) AC Hall effect measurement for PBTTT:TFSI (100/1 mM BMP-TFSI / FeCl$_3$, 100 sec); device conductivity was 900 S/cm. f) Hall bar structure used for measurement.}
\label{n}
\end{center}
\end{figure*}

\begin{table*}[t]
\begin{center}
\onehalfspacing
\caption{Carrier densities measured by XPS and QNMR} \label{Ntab}
\begin{tabular}{@{}llllll@{}}
\toprule
       & & \multicolumn{2}{c}{S 2p XPS} & \multicolumn{2}{c}{$^{19}$F QNMR}                      \\
Polymer:Ion & Doping time & Molar doping ratio & N (10$^{20}$ cm$^{-3}$) & Molar doping ratio & N (10$^{20}$ cm$^{-3}$)              \\
\midrule
PBTTT:TFSI & 100 s  & 0.851 $\pm$ 0.023  & 7.71 $\pm$ 0.21 	& 0.822 $\pm$ 0.074  & 7.45 $\pm$ 0.67  \\
PBTTT:TFSI & 300 s  & 0.967 $\pm$ 0.020  & 8.77 $\pm$ 0.18	& 0.916 $\pm$ 0.071  & 8.31 $\pm$ 0.64   \\
P3HT:HFSI & 300 s & 0.307 $\pm$ 0.042 & 12.7 $\pm$ 1.7 & -- & -- \\
DPP-BTz:HFSI & 300 s & 0.588 $\pm$ 0.064 & 3.05 $\pm$ 0.33 & -- & -- \\
IDTBT:HFSI & 300 s & 0.787 $\pm$ 0.028  & 3.65 $\pm$ 0.13 & -- & -- \\
\bottomrule
\end{tabular}
\end{center}
\end{table*}

XPS allows us to characterize the atomic species present near the film surface, because the ratio of peak integrals for a given transition are proportional to the molar ratio of atoms in each species. \cite{Moulder:1992,Nefedov:1988} Both TFSI and PBTTT contain sulfur, therefore by measuring the areas of the sulfur peaks corresponding to the ion and polymer we can determine the molar ratio of TFSI to PBTTT. XPS is highly surface sensitive, so this measurement is only possible when the surface of the polymer is clean (i.e. contains no residual ionic liquid, which would skew the ratio of TFSI to PBTTT). Our doping process includes a surface washing step which eliminates any surface residue, which we additionally confirm by NMR (Figure \ref{n}d; further details in Supporting Information Section 2).

Figure \ref{n}a-c shows the XPS sulfur 2p spectra of three samples, along with fits to the experimental spectra. First, we fit the spectrum of a sample doped by FeCl$_3$ without ion exchange (Figure \ref{n}a) to allow us to assign the spectral features originating from the polymer. We were unable to obtain good fits using a single S 2p doublet, however a fit using two doublets constrained to a 1:1 area ratio shows excellent agreement with the measured signal. The 1:1 ratio of these two S doublets suggests these signals derive from the inequivalent thiophene and thienothiophene sulfur atoms in PBTTT.

Figure \ref{n}b shows fits to an ion exchange doped PBTTT:TFSI film (100/1 mM BMP TFSI / FeCl$_3$, 100 sec exposure). Here, we see the appearance of a new set of bands, chemically shifted to higher binding energy, consistent with the increased electron density of TFSI. We again are unable to fit this band using a single S 2p doublet, but were able to achieve a good fit by again adding a second S 2p doublet, again constrained by a 1:1 area ratio. This observation suggests an inequivalent chemical environment of the two sulfur atoms in TFSI, consistent with the structural distortion of TFSI obtained from atomistic modeling discussed later (Figure \ref{pbttt}c). Although the resulting model, consisting of 4 S 2p doublets (i.e. 8 total Voigt peaks) appears complex, the fit has only five key parameters: the position of the 2p$_{3/2}$ peak of each of the four doublets, and the ratio of the PBTTT to TFSI peak areas. The same model provides an equally good fit to the 300 s doped sample (Figure \ref{n}c). The mole ratio of PBTTT to TFSI can be directly extracted from the PBTTT to TFSI peak area in each fit, scaled by a factor of 2 to account for the different number S atoms in the PBTTT monomer vs. TFSI. The resulting molar ratios approach 1 ion per monomer in PBTTT:TFSI at 300s; molar ratios and corresponding carrier densities are given in Table \ref{Ntab}. Further discussion and XPS spectra for P3HT, DPP-BTz, and IDTBT are given in Supporting Information Section 2.

We further validate these measurements using qantitative nuclear magnetic resonance (QNMR),  another straightforward method for measuring the absolute concentration of a chemical species. In a QNMR experiment, the integrated NMR signal from the species of interest is compared with that of an internal standard of known concentration. As long as certain conditions are met\textemdash effectively ensuring that the system returns fully to thermal equilibrium between each scan\textemdash the integrated intensity ratios of each signal in an NMR spectrum is proportional to the mole ratio of the corresponding nuclei.\cite{Holzgrabe:2010} While in principle it is possible to directly measure the concentration ratio of the polymer and ion without a standard, this is not straightforward in PBTTT:TFSI because the only nuclei shared by both PBTTT and TFSI is $^{33}$S, which has low natural abundance and is relatively insensitive. $^{19}$F, on the other hand, is nearly 100\% natural abundant and only slightly less sensitive than $^1$H, making it ideal for quantifying relatively small amounts of material. This factor is critical since even at a 1:1 mole ratio relative to PBTTT monomers, in thin films there are only a few nmol of TFSI per cm$^2$ of film area.

Solution-state NMR is typically much more sensitive than solid-state NMR because motional averaging due to molecular tumbling in solution results in extremely narrow linewidths. To take advantage of this sensitivity boost, we extract the TFSI ions from of doped thin films using a chemical dedoping process,\cite{Jacobs:2017b} then measure the NMR spectrum of the dedoping solution along with a known amount of a QNMR reference compound, 4,4-difluorobenzophenone (DFBP), using high-resolution solution state NMR. UV-vis spectra were collected after dedoping the films (Figure S14) revealing nearly complete dedoping, consistent with previous reports. A similar approach was previously used to quantify the concentration of dilute co-solvents in polymer thin films, allowing detection down to less than 1:1000 relative to P3HT monomers.\cite{Jacobs:2018a,Chang:2013} 

Figure \ref{n}d shows the $^{19}$F QNMR spectra; the carrier densities are obtained by dividing the TFSI concentration obtained from integrating the NMR peak by the film volume. We see excellent agreement with the corresponding values obtained from XPS (Table \ref{Ntab}); both samples are consistent to well within the error bounds of each measurement. 

In contrast, the Hall effect does not give a reliable measurement of carrier density in conducting polymers, despite its occasional use in the literature for this purpose. Figure \ref{n}e shows AC Hall effect measurements of an ion exchange doped PBTTT:TFSI device (structure shown in Figure \ref{n}f). The carrier density obtained from this measurement is $4.2 \times 10^{21}$ cm$^{-3}$, about 5 times larger than the values measured by XPS and QNMR. Although the origin of this deviation is not completely clear, in general the Hall coefficient, and thus the carrier density, shows a significant temperature dependence which is believed to result from partial screening of the Hall voltage by hopping carriers.\cite{Yi:2016} In the formalism given by Yi \textit{et al.}, they consider two populations of charge carriers, corresponding to a population of fully delocalized states which contribute to the Hall voltage, and another set of fully localized states which do not generate a Hall voltage but still move in response to the Hall voltage. This screening effect leads to an underdeveloped of the Hall voltage, which in turn yields an overestimate of the carrier density.

\subsection*{Ionic size effects on structural disorder and conductivity}

\begin{figure*}[t!]
\begin{center}
\includegraphics[width = \textwidth]{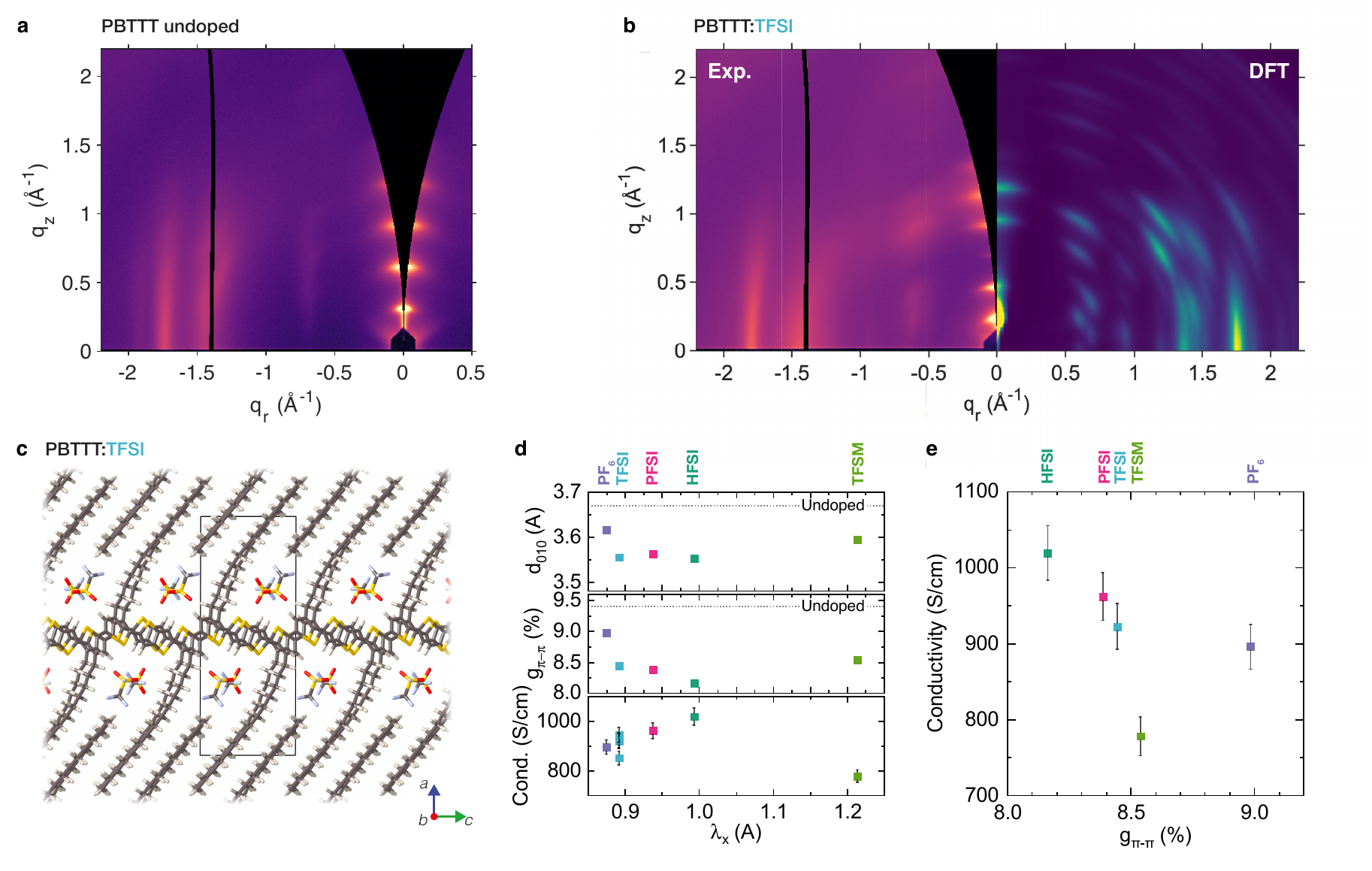}
\caption{\textbf{Ionic size effects in PBTTT}. a) Experimental GIWAXS pattern for undoped PBTTT. b) Experimental and simulated GIWAXS patterns for PBTTT:TFSI. c) Optimized structure PBTTT:TFSI at 1:1 molar doping level.  d) Plot of $\pi-\pi$ stacking distance (top), $\pi-\pi$ paracrystallinity (middle), and electrical conductivity (bottom) vs. the smallest principal moment of the ionic gyration tensor, $\lambda_x$. e) Plot of electrical conductivity vs $\pi-\pi$ paracrystallinity.}
\label{pbttt}
\end{center}
\end{figure*}

We selected PBTTT:TFSI as a model system to better understand the atomic scale packing and simulated several possible packing motifs and generated GIWAXS patterns for each.\cite{Lemaur:2013} These simulations confirm that TFSI packs between the alkyl chains and in close contact with the PBTTT conjugated cores, (Figure \ref{pbttt}c), as previously proposed for F4TCNQ\cite{Kang:2016} and fullerenes.\cite{Mayer:2009} Experimental GIWAXS patterns of undoped PBTTT (Figure \ref{pbttt}a) and PBTTT:TFSI (Figure \ref{pbttt}b) show significant changes to the polymer crystal structure upon doping, including a significant expansion of the lamellar stacking (100) distance from 20.2 \AA \ to 26.5 \AA, a slight contraction of the $\pi$-stacking (010) distance, and a reduction in intensity of the (300) reflection intensity. Figure \ref{pbttt}b shows the GIWAXS pattern generated from the simulated 1:1 molar doping level structure, consistent with the experimental carrier density measurements in Figure \ref{n}. We observe good qualitative agreement with experiment, in particular reproducing the mixed-index peaks at $q_y$ = 0.6 \AA$^{-1}$ and the reduced intensity of the (300) and (600) peaks, as well as nearly quantitative agreement with the experimental unit cell parameters (Supporting Information Section 4), which showed some sample-to-sample variation in our previous study.\cite{Jacobs:2021a} Similar diffraction pattern features, including strongly reduced (300) intensity and peaks at $q_y$ = 0.6 \AA$^{-1}$ are also observed in the HFSI and PFSI experimental GIWAXS patterns (Supporting Information Section 7.1), suggesting a similar packing motif for these three ions.

The low dielectric constant of most organic semiconductors (typically $\epsilon_r \sim 3$) causes dopant ions to generate local potential wells with depths $\gg kT$ ($\sim$100s of meV).\cite{Arkhipov:2005} The depth of these wells are set by the minimum approach distance of the dopant ion and the charged polymer backbones as well as the distance between dopants. Many of the ions studied here are non-spherical, complicating the analysis of ionic size. However, given the strength of the hole-ion Coulombic interaction ($\gg kT$), we expect that in an ICTC, non-spherical dopant ions should orient themselves to minimize the center of charge distance between themselves and the hole, maximizing their Colombic stabilization. Under this assumption, the ion-polaron distance is proportional to the smallest principal component of the ion gyration tensor, $\lambda_x$, which describes the smallest semi-axis of an ellipsoid representing the shape of the molecular ion. Therefore, if ionic trapping dominates the conductivity, we should observe an increase in conductivity with increasing $\lambda_x$. Although this simple approach accounts only for the leading monopole interactions and neglects polarizability effects, it should be sufficient to reveal qualitative trends.

Figure \ref{pbttt}e shows the conductivity of PBTTT films doped with each ion plotted vs. $\lambda_x$. We observe a modest increase in conductivity with increasing $\lambda_x$, which then reverses for the largest ion, TFSM. The magnitude of the effect is quite small, however. We also tried to correlate the conductivity with the $\pi-\pi$-stacking distance, which has previously been shown to govern transport in conjugated polymers\cite{Noriega:2013}; a decrease in $d_{010}$ should increase the hopping transfer integral and conductivity. Although the $\pi-\pi$-stacking distance $d_{010}$ (Figure \ref{pbttt}e) is higher for PF$_6$, the values are similar for TFSI, PFSI, and HFSI. Therefore, a decrease in $d_{010}$ cannot explain the increased conductivity from TFSI to HFSI.

We also extracted the $\pi$-stacking paracrystallinity for PBTTT doped with each ion from the GIWAXS data. Paracrystallinity is a measure of cumulative disorder in a crystal which originates from a statistical variation in stacking distances. The paracrystallinity parameter $g_{\pi-\pi}$ quantifies the magnitude of this disorder as the standard deviation in stacking distance normalized by the stacking distance, i.e. $g_{\pi-\pi} = \delta_{\pi-\pi} / d_{\pi-\pi}$. Rivnay \textit{et al.} previously demonstrated that paracrystallinity is typically the dominant peak broadening mechanism in conjugated polymer GIWAXS data.\cite{Rivnay:2011a} As seen in Figure \ref{pbttt}e, paracrystalline disorder and ionic size are not independent; rather paracrystallinity drops with respect to undoped PBTTT (indicated by dotted line) for all ions measured. We attribute this effect to increasing 2d polaron delocalization and increased backbone planarity, which will reduce both $d_{\pi-\pi}$ and $g_{\pi-\pi}$.\cite{Liu:2018a,Kang:2016} This effect becomes stronger with increasing $\lambda_x$, presumably due to a decrease in electrostatic disorder; in this context the trend reversal for TFSM may be due to the lower ion exchange efficiency for large ions in PBTTT (Supporting Information Section 1). The reduction in disorder with increasing ionic size seen experimentally should also contribute to the observed variation in conductivity between ions. Indeed, Figure \ref{pbttt}f shows a plot of conductivity vs. g$_{\pi-\pi}$, revealing a clear trend of increasing conductivity as structural disorder decreases. However, due to the small changes in conductivity and g$_{\pi-\pi}$ observed here, it is difficult to conclude based on measurements of PBTTT alone whether the observed variation in conductivity originates primarily from ionic trapping or structural disorder. However, it is clear that in PBTTT, high crystallinity along with significant free volume in the lamellar stacking region allow for low structural disorder even at very high doping levels. In this regime, we see no clear evidence for reduced ICTC binding energy with increasing dopant size, suggesting that if ICTCs do still behave as traps at this doping level, they must be very shallow.

\begin{figure*}[t!]
\begin{center}
\includegraphics[width = \textwidth]{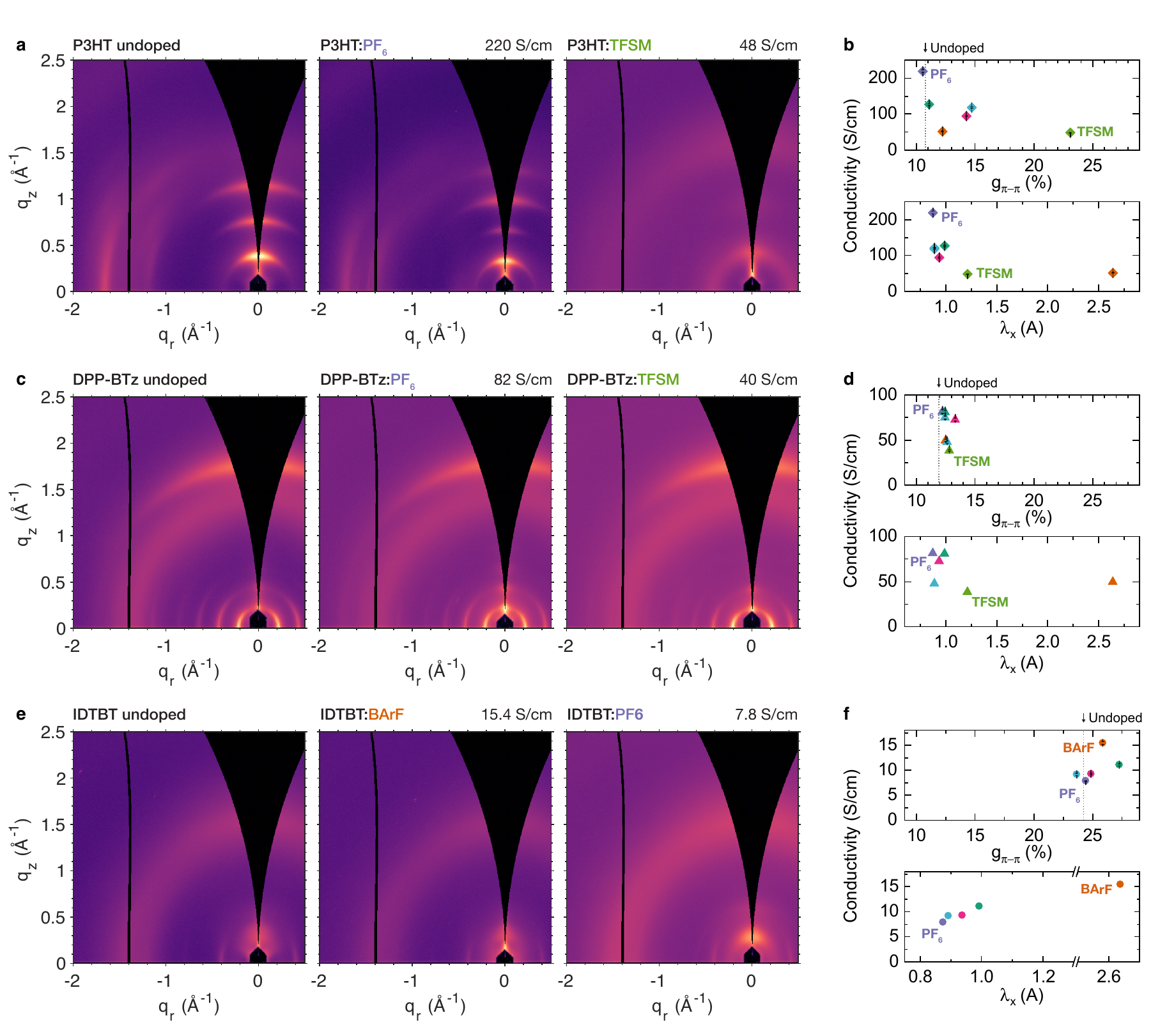}
\caption{\textbf{Ionic size effects in P3HT, DPP-BTz, and IDTBT}. GIWAXS patterns for P3HT (a), DPP-BTz (c), and IDTBT (e), showing the undoped material (left) and each polymer doped with the ion giving the highest (center) and lowest (right) conductivity. Doping solution exposure time is 5 minutes, corresponding to saturation doping level. Plot of electrical conductivity vs $\pi$-stacking paracrystallinity (top), and electrical conductivity vs. the smallest principal moment of the ionic gyration tensor, $\lambda_x$ (bottom) for P3HT (b) DPP-BTz (d) and IDTBT (f). For P3HT and DPP-BTz the conductivity for 5 min doping time is shown; values for IDTBT corresponds to the highest conductivity measured for each ion (see Supporting Information Section 3)}
\label{others}
\end{center}
\end{figure*}

PBTTT is an unusually crystalline material, so to put these findings in a more general context, we compared the effect of ion size and doped film paracrystallinity on conductivity in three other polymers with microstructures ranging from polycrystalline to highly disordered. Figures \ref{others}a,c,e show GIWAXS patterns for each polymer before doping (left), and after 5 minute ion exchange doping with the ion yielding the highest and lowest conductivity (center and right panels, respectively). 

In the polycrystalline polymer P3HT (Figure \ref{others}a), we observe dramatic variation in crystallinity upon doping. The smallest ion studied here, PF$_6$, yields a highly crystalline film with a slight increase in lamellar stacking distance and decreased $\pi$-stacking distance and paracrystallinity, consistent with our observations in PBTTT. These results suggest incorporation of the PF$_6$ ion into the sidechain region, similar to the structures proposed for P3HT:F4TCNQ.\cite{Hamidi-Sakr:2017,Scholes:2017} Doping with TFSM, on the other hand, yields a nearly amorphous film, with only the (100) peak ($q_z = 0.4$\AA$^{-1}$) and a broad $\pi$-stacking halo ($q = 1.5-1.6$\AA$^{-1}$) still discernible. Intriguingly, the lamellar stacking distance observed for TFSM is shorter than undoped P3HT, similar to the metastable fractional CT phase of P3HT:F4TCNQ,\cite{Jacobs:2018,Stanfield:2021} suggesting that TFSM may be intercalating into the polymer $\pi$-stacks. For further discussion, see Supporting Information Section 1.1. Figure \ref{others}b shows the conductivity for P3HT films doped with each ion plotted vs. paracrystallinity (top) and $\lambda_x$ (bottom). We observe a clear increase in conductivity with decreasing paracrystallinity, as expected, reaching a maximum value of 220 S/cm for PF$_6$. This is an exceptionally high value for P3HT, over two orders of magnitude higher than typically achieved with molecular dopants such as F4TCNQ, and matching a recent report for electrochemically prepared P3HT:PF$_6$.\cite{Neusser:2020} Remarkably, this high value is only achieved using the smallest ion; as the ionic size is increased, conductivity \textit{decreases}, in stark contrast with the predictions of the Arkhipov model.\cite{Arkhipov:2005} Therefore, in P3HT the effect of the ion on the polymer microstructural order appears to be more important than its Colombic interaction with charge carriers.

In contrast, DPP-BTz, a high mobility donor-acceptor co-polymer with moderate crystallinity, displays relatively little variation in the diffraction pattern between ions (Figure \ref{others}b). Lamellar stacking distances increase by less than 1 \AA~  upon doping while $\pi$-stacking distances are consistent to within 0.05 \AA~ (Supporting Information Section 4.3). Nonetheless, we still observe the same trends with respect to conductivity as seen in P3HT\textemdash conductivity is again inversely correlated with ion size, reaching 80 S/cm for PF$_6$, and improves with decreasing paracrystallinity (Figure \ref{others}d). Again, this unexpected result suggests ionic trapping is negligible, since the small microstructural changes observed from GIWAXS should amplify the effect of any ionic size on conductivity. Instead, our results indicate that in DPP-BTz, even quite small variations in microstructure are more important to charge transport than ion size.

Only IDTBT, the most disordered material studied here, shows qualitatively different behavior from the other polymers. As with DPP-BTz, we observe little change in the GIWAXS pattern upon doping; each of our IDTBT GIWAXS data (Figure \ref{others}e) show a broad out-of-plane $\pi$-stacking peak at $q_z = 1.6$ \AA$^{-1}$ and a lamellar stacking peak at $q_z \approx 0.4$ \AA$^{-1}$, along with a sharp in-plane diffraction peak at $q_r = 0.2$ \AA$^{-1}$ assigned to the backbone repeat stacking (001).\cite{Zhang:2013} Despite its high FET mobility and extremely low energetic disorder,\cite{Venkateshvaran:2014} the highest conductivity achieved in IDTBT is quite low, reaching only 15.4 S/cm (Figure \ref{others}f). Conductivity in IDTBT was also found to be highly sensitive to doping level, decreasing significantly at high doping levels, in contrast with the other three polymers (see Supporing Information Section 3). Conductivities reported for IDTBT therefore correspond to the peak conductivity measured for each ion. Furthermore, we observe no clear correlation between $\pi$-stacking paracrystallinity and conductivity in IDTBT; instead, conductivity steadily increases with increasing ionic size. Although we cannot confidently explain these observations, a plausible theory for the behavior of IDTBT is given below.

\subsection*{Correlation between paracrystalline disorder and conductivity}

\begin{figure}
\begin{center}
\includegraphics[width = \columnwidth]{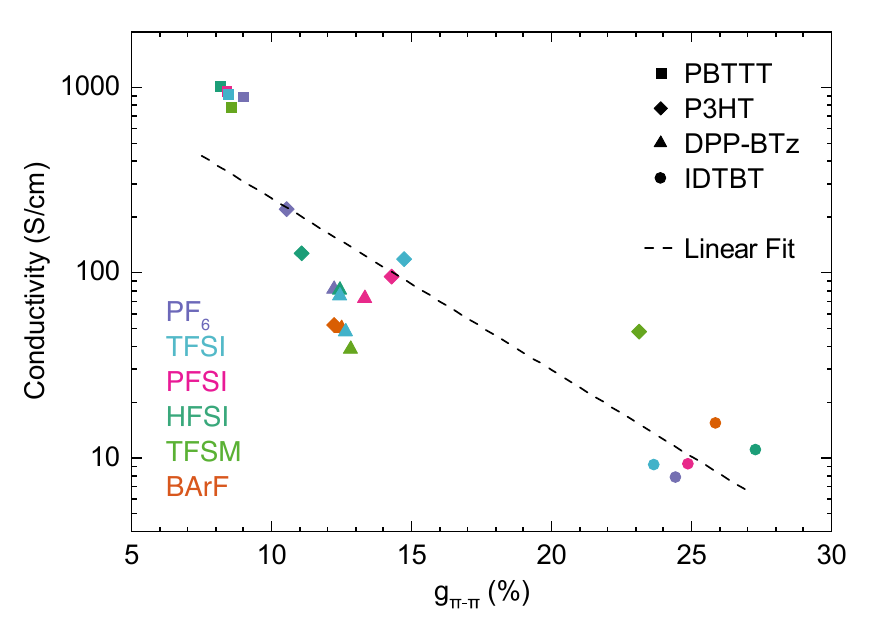}
\caption{\textbf{Effect of paracrystalline disorder on conductivity}. Conductivity vs. $\pi$-stacking paracrystallinity for four different polymers doped with different ions. Symbol color corresponds to dopant ion; symbol shape corresponds to polymer (structures given in Figure \ref{Fig1}d,e. Dashed line indicates linear fit to the full dataset with slope $d\log_{10}{\sigma}/dg_{\pi-\pi} = - 9.3 \pm 1.2$}
\label{g}
\end{center}
\end{figure}

Figure \ref{g}a shows the conductivity of all four polymers doped with various ions plotted together vs. $\pi-\pi$ paracrystallinity, revealing an unexpectedly strong correlation between these two quantities. The strength of this correlation is surprising, particularly given that IDTBT and DPP-BTz exhibit up to an order of magnitude higher FET mobility than P3HT and PBTTT, which by the relation $\sigma = en\mu$ one might expect to see reflected in a higher, rather than lower conductivity. A similar plot of conductivity vs. $\lambda_x$ (Figure S26) shows no correlation. This observation suggests that, at least at high doping levels relevant to many device applications, \textit{the most important factor in achieving good charge transport is not minimization of ionic trapping, as previously assumed, but maximization of structural order}.

An important implication of this finding is that doping efficiencies in PBTTT, P3HT, and DPP-BTz is almost certainly near 100\%. Several other observations also support this idea. First, the highest conductivities in P3HT and DPP-BTz are both achieved upon doping with PF$_6$, the smallest of the ions studied here, which should give the lowest doping efficiency if trapping was the dominant effect. Additionally, if we assume all dopant ions generate a free charge carrier, we can estimate a lower bound on the carrier mobilities: for PBTTT:TFSI $\mu_h \geq 8$ cm$^{2}$ V$^{-1}$ s$^{-1}$, while for P3HT:PF6 $\mu_h \geq 1$ cm$^{2}$ V$^{-1}$ s$^{-1}$. These values are both over an order of magnitude higher than the FET mobility of the undoped polymers;\cite{McCulloch:2006,Wang:2010} lower doping efficiency would imply even higher carrier mobilities.

\begin{figure*}[ht]
\centering
\includegraphics[width=\textwidth]{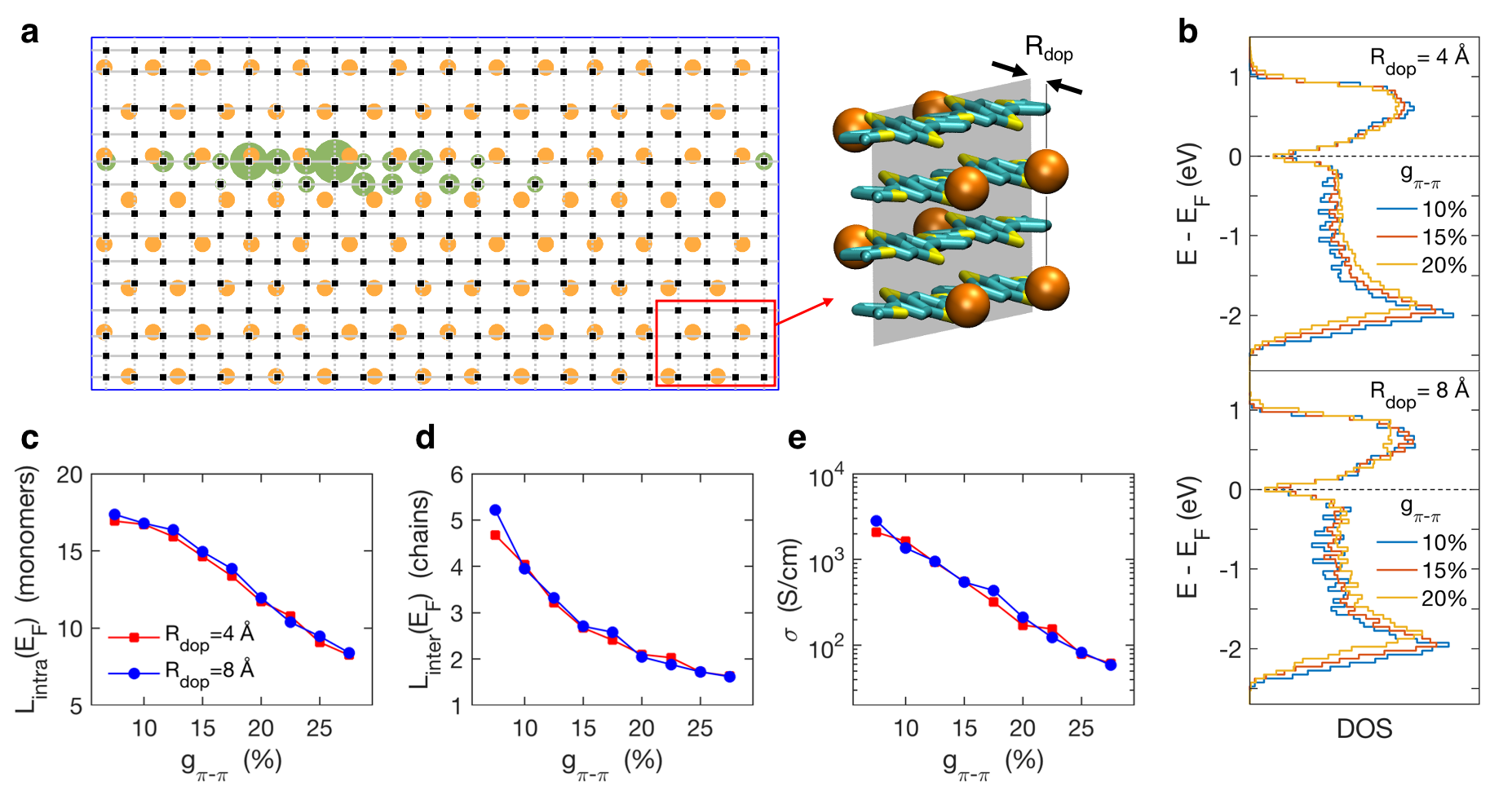}
\caption{\textbf{Theoretical model for highly-doped polymers.} 
(a) Sketch of the 2D lattice of a doped polymer lamella characterized by paracrystallinity $g_{\pi-\pi}=20$\%. 
As shown in the atomistic model, black squared and orange dots correspond to monomer sites and dopant ions, respectively. 
The green-shaded region depicts the hole density for a typical localized state at $E_F$.
(b) Density of states as a function of paracrystallinity and dopant-ions distance, displaying a negligible dependence on these parameters.
Paracrystallinity and dopant distance dependence of intra-chain (c) and inter-chain localisation length (d) in units of monomers and chains, respectively, and conductivity (e). Model results elucidate the degradation of transport properties with paracrystalline disorder and the negligible impact of the ion size.
}
\label{theory}
\end{figure*}

\subsection*{Theoretical transport modeling}

Our experimental analysis points to the primary role of paracrystallinity in controlling charge-transport properties of semi-crystalline polymers doped by ion-exchange. 
To rationalize this intriguing observation, we propose a general model for the electronic structure of heavily-doped polymers. Our model encompasses paracrystalline disorder and long-range Coulomb interactions among holes on the polymer chains and with the ions. As shown in Figure~\ref{theory}a, we model a paracrystalline lamella of a polymer such as PBTTT as a 2D lattice with irregular spacing along the $\pi$-stacking direction. 
Ions are placed at distance $R_\mathrm{dop}$ above and below the plane of the $\pi$-backbones, corresponding to incorporation in the alkyl chains region. 
Atomistic calculations enabled a careful parameterization of the model, including the quantification of the energetic disorder arising from paracrystallinity. 
Indeed, the structural paracrystalline disorder determines an increase with $g_{\pi-\pi}$ of both the local and non-local energetic disorder experienced by the holes on the polymer chains. 
Though specific to PBTTT, our model parameterization is broadly representative of the entire set of semi-crystalline polymers considered in this study. Our model will be used to rationalize the general trends as a function of paracrystallinity and $R_\mathrm{dop}$, the latter parameter mimicking the size of molecular ions.

Our calculations reveal that, for all paracrystallinity values and dopant-polymer distances considered, the density of states (DOS, Figure~\ref{theory}b) is characterized by a dip at the Fermi level ($E_F$), a result that is consistent with photoemission data on PEDOT-PSS.\cite{Bubnova2014}
This is the signature of a Coulomb gap originating from hole-hole interactions at large charge density,\cite{Efros1975}
which largely suppress the number of states available for transport. 
In addition, states at the Fermi level, i.e. those contributing to charge transport, are significantly more localized (less mobile) than deeper occupied or shallower unoccupied states, both in terms of spatial extension of the their wavefunctions along polymers chains and between multiple chains.
The intra-chain and inter-chain delocalization of the states at $E_F$ decrease with paracrystallinity, as shown in Figure~\ref{theory}c,d, which reveals the effect of the ensuing energetic disorder in the electronic states. 
The occurrence of a Coulomb gap at the Fermi level has been reported in recent kinetic Monte Carlo simulations,\cite{Koopmans:2020,Fediai:2019} which, hinging on the assumption of charges localized on molecular units, could not grasp the effect of electron-electron interactions in limiting the size of carriers wave packets. 

The localized nature of electronic states at $E_F$, together with the dynamic nature of energetic disorder in soft organic materials, enables us to compute the dc electrical conductivity in the framework of the transient localization theory.\cite{Fratini16,Fratini:2020} 
The starting point is the Kubo-Greenwood formalism, within which the  conductivity for a two-dimensional system with purely static disorder would be strictly zero.\cite{Lee:1985} However, the disorder is never static in soft organic materials, because it is modulated by low-frequency thermal lattice vibrations, conferring a finite diffusivity to charge carriers. We use the relaxation time approximation to account for these TL phenomena. We note that while previous implementations of the relaxation time approximation implicitly considered all energetic disorder as dynamic, as appropriate in undoped molecular crystals, 
it is not \textit{a priori} clear whether this is strictly the case in heavily doped polymers. Nonetheless, this phenomenological theory should apply also to the case where part of the disorder is static,\cite{Lee:1985} as further corroborated by the excellent agreement with experimental data discussed below.

Calculation results in Figure~\ref{theory}e show that paracrystalline disorder is the leading factor determining the two-order of magnitude drop in the conductivity with paracrystallinity. Besides capturing the correct order of magnitude for conductivity, the theory predicts an exponential suppression of $\sigma$ upon increasing $g_{\pi-\pi}$, with decay rate $d\log_{10}{\sigma}/dg_{\pi-\pi} = - 8.4 \pm 0.3$; this is in excellent agreement with
the best fit to the experimental data in Figure~\ref{g}, which gives $d\log_{10}{\sigma}/dg_{\pi-\pi} = - 9.3 \pm 1.2$. 
Importantly, our calculations also reveal the negligible role of the ion size in determining the conductivity of the material.

Multiple factors conspire to eliminate the effects ionic trapping at high doping levels. The first of these, the smoothing of the Coulomb landscape at large ion density,\cite{Arkhipov:2005} is well known. However, our calculations reveal two additional contributions that were  previously not well appreciated: (i) the further smoothing of the Coulomb landscape by repulsive hole-hole interactions, which partially neutralize the attractive interactions with dopant ions and (ii) the large spatial extension of the electronic wavefunctions probing this energy landscape (see Figure~\ref{theory}a,c,d), preventing localization and trapping of charge carriers within a single Coulomb well.
Our theoretical results hence rationalize the leading role of paracrystallinity in being the most critical parameter, among many others in the complex transport physics, in controlling the charge transport properties of these ion-exchanged doped polymers at high doping levels.

\section*{Discussion and conclusions}

Our work has demonstrated that ion-exchange doping with FeCl$_3$ can generate highly ordered polymer films with extremely high doping levels approaching one ion per monomer and very high conductivities above 1000 S/cm. Our combined experimental and theoretical results demonstrate that in this high doping regime, relevant to most device applications, enhanced crystallinity is the most critical factor for achieving high conductivities. 

The irrelevance of ionic trapping is demonstrated by the negligible correlation of conductivity with the ion size in all the polycrystalline polymers studied here. The only exception we observe is IDTBT, which displays a highly disorder microstructure in GIWAXS (Figure \ref{others}e) but a highly planar backbone and extremely low electronic disorder.\cite{Venkateshvaran:2014} Ion-exchange doped IDTBT displays several unique features, including a strong correlation between $\lambda_x$ and conductivity (Figure \ref{g}b) and a strong decrease in conductivity at high doping levels. We cannot expect the two-dimensional, lamellar microstructure assumed in our model to accurately describe near-amorphous systems like IDTBT. 

We propose two factors which may contribute to the unique behavior of IDTBT. First, interchain transport in IDTBT is believed to be primarily mediated by close contacts between BT groups at crossings between only two chains,\cite{Thomas:2019,Cendra:2021} rather than through longer-range interchain delocalization in larger aggregates, as in most other polymers.\cite{Noriega:2013} We expect that the localized nature of these interchain crossing states\cite{Thomas:2019} should make them more suceptible to trapping by nearby ions, in contrast with the more delocalized in larger $\pi$-aggregates in polycrystalline systems. Second, it is likely that in IDTBT additional space is provided by the disordered microstructure, due to the long sidechains that extend beyond the backbone plane and prevent close $\pi$-$\pi$ stacking. This microstructure may allow for more intimate contact between the polymer backbone and the dopant ion. The combination of these two factors could lead to a strongly ion-size dependent electron transfer rate at localized chain crossings sites. We hypothesize that the presence of an ion near a chain crossing could energetically shift the two crossing sites out of resonance, increasing the activation energy. Such a mechanism would be suppressed in larger aggregates or in materials with a higher density of chain crossings, suggesting that the unique microstructure of IDTBT, recently revealed via TEM\cite{Cendra:2021} may be key to understanding these effects.

The very good quantitative agreement between experiment and theory both in terms of the magnitude of the electrical conductivities and the dependence of conductivity on paracrystallinity suggests that TL provides a powerful framework for understanding the charge transport properties of highly doped conducting polymers.  This allows identifying new strategies for future optimization of doped polymers, including through further reductions in paracrystallinity. Our calculations also suggest that in current systems achievable conductivities are partly limited by the suppression of the density of states and localisation length near the Fermi level, caused by the Coulomb repulsion between the carriers. This mechanism is predicted to be very sensitive to the polymer reorganisation energy related to the high-frequency intramolecular vibrations  and to intra-chain charge-hopping couplings  (Figure S35). By reducing reorganisation energy and/or increasing the intramolecular interactions it might be possible to completely suppress the Coulomb gap and enter a truly metallic regime with significantly higher conductivities.

\section*{Acknowledgements}
I.E.J acknowledges funding through a Royal Society Newton International Fellowship. Financial support from the European Research Council for a Synergy grant SC2 (no. 610115) and from the Engineering and Physical Sciences Research Council (EP/R031894/1) is gratefully acknowledged. Y.L. thanks the European Commission for a Marie-Sklodowska-Curie fellowship. For PhD fellowships D.S. thanks the EPSRC CDT in Sensor Technologies for a Healthy and Sustainable Future (Grant No. EP/L015889/1), W.W. the EPSRC CDT in Connected Electronic and Photonic Systems, L.J.S the ERC Synergy Grant SC2 (Grant No. 610115), T.M. the EPSRC CDT in Nanoscience and Nanotechnology, L.L. the EPSRC CDT in graphene technology and D.T. the Cambridge Commonwealth European and International Trust. S.B. and S.R.M. thank National Science Foundation (through the DMREF program, DMR-1729737). This research used resources of the Advanced Photon Source, a U.S. Department of Energy (DOE) Office of Science User Facility, operated for the DOE Office of Science by Argonne National Laboratory under Contract No. DE-AC02-06CH11357. We thank Yadong Zhang for some dopant synthesis, and Carmen Fernandez Posada and Mohamed Al-Hada for assistance with XPS measurements.

\section*{Methods}

\subsection*{Materials}
PBTTT (poly(2,5-bis(3-alkylthiophen-2-yl)thieno(3,2-b)thiophene); Mw = 44 kDa, PDI = 1.47), IDTBT-C16 (poly(indaceno(1,2-b:5,6-b')dithiophene-co-2,1,3-benzothiadiazole); (Mw = 92 kDa, PDI 2.3), and DPP-BTz (poly((2,5‐bis(2‐octadecyl)‐2,3,5,6‐tetrahydro‐3,6‐diketopyrrolo(3,4‐c)pyrrole‐1,4‐diyl)‐alt‐(2‐octylnonyl)‐2,1,3‐benzotriazole); Mw = 63 kDa, PDI = 3.2) were synthesized as described previously.\cite{McCulloch:2006,Zhang:2010,Gruber:2015} P3HT (poly(3‐hexylthiophene‐2,5‐diyl); 99.0\% RR, Mw = 44 kDa, PDI 2.1) was purchased from TCI. Ion-exchange salts Li-PFSI (\textgreater 98\%), Li-HFSI (\textgreater 98\%), and Na-BArF  (\textgreater 98\%, \textless 7\% water) were purchased from TCI; Li-TFSI (\textgreater 99\%, \textless 1\% water), Na-TFSI (\textgreater 97\%), BMP-TFSI (\textgreater 98.5\%, \textless 0.04\% water), EMIM-TFSI (\textgreater 98\%, \textless 0.1\% water), TBA-TFSI (\textgreater 99\%), DMPI-TFSM (\textgreater 97\%, \textless 0.5\% water), TBA-OTf (\textgreater 99\%), and TBA PF6 (\textgreater 99\%) were purchased from Sigma Aldrich. Dopants PMA (hydrated, ACS reagent), Fc-PF$_6$ (\textgreater 97\%), Cu(OTf)$_2$ (\textgreater 98\%), FeCl$_3$ (anhydrous, \textgreater 99.99\% trace metals basis), OA, and CAN (\textgreater 99.99\% trace metals basis) were purchased from Sigma Aldrich. F4TCNQ (\textgreater 98\%) was obtained from TCI. TBA CN6-CP, F6TCNNQ, Mo(tfd)$_3$, Mo(tfd-COCF$_3$)$_3$, and CN6-CP were synthesized as described previously\cite{Karpov:2018,Koech:2010,Davison:1964,Paniagua:2014,Fukunaga:1976,Karpov:2016}. All polymer and dopant solutions were prepared using anhydrous solvents (Romil Hi-Dry, \textless 20 ppm water). Triethylamine (\textgreater 99.5\%) and 4,4'-difluorobenzophenone (TraceCERT certified reference material) for QNMR dedoping experiments were obtained from Sigma Aldrich. All materials were used as received with the exception of Na BArF, which was dried following the procedure given by Yakelis \textit{et al.}\cite{Yakelis:2005}

\subsection*{Solution Preparation}
PBTTT, P3HT, and IDTBT solutions (10 mg/mL, 1,2-dichlorobenzene (DCB)) and heated at 80$^{\circ}$ C overnight before use. DPP-BTz solutions were prepared at the same concentration in chlorobenzene and heated at 110$^{\circ}$ C following the procedure in Schott \textit{et al.}\cite{Schott:2015} Electrolyte solutions for ion exchange doping were prepared at 1M concentration in acetonitrile, FeCl$_3$ solutions were prepared at 10 mM concentration. Electrolyte solutions remained stable in the glovebox for extended periods, however FeCl$_3$ solutions were always prepared immediately before use. All solution preparation and reagent weighing was performed under nitrogen atmosphere (\textless 1 ppm H$_2$O, O$_2$ during solution preparation; \textless 10 ppm H$_2$O, O$_2$ during weighing). 

\subsection*{Sample Preparation}
Electrical conductivity and UV-vis were measured on 1 cm square glass substrates (Corning Eagle XG) with 1 mm van der Pauw contacts covering each corner (thermally evaporated Cr/Au, 5/25 nm). GIWAXS samples were coated on 1.5 cm square bare Si substrates. All substrates were cleaned by sequential sonication in 2\% Decon 90/DI water, DI water, acetone, and isopropanol, dried by nitrogen gun, then etched with oxygen plasma (300 watts, 10 minutes) before use. 

PBTTT films were spin coated at 1500 rpm for 60 seconds from 80$^{\circ}$ C solution, using glass pipettes and substrates preheated to the same temperature. IDTBT and P3HT films were spin coated from 60$^{\circ}$ C solutions using the same procedure. DPP-BTz were spun from 110$^{\circ}$ C solutions at 2000 rpm.

PBTTT and P3HT samples were subsequently annealed in N$_2$ at 180$^{\circ}$ C for 20 minutes, then slowly cooled to room temperature by switching off the hotplate. IDTBT samples were dried at 100$^{\circ}$ C for 5 minutes after spin coating. DPP-BTz films were annealed at 110$^{\circ}$ C for 1 hour then quenched following the procedure of Schott \textit{et al.}\cite{Schott:2015}

Ion-exchange doping was performed following the procedure published previously.\cite{Jacobs:2021a} Briefly, films were sequentially doped on the spin coater by covering the sample with an electrolyte/FeCl$_3$ solutions (100mM/1mM in acetonitrile, unless otherwise specified), waiting for a variable delay period (300 s unless otherwise specified), then spinning off the excess. While the sample is still spinning, the doped film was washed with 1 mL acetonitrile to remove any electrolyte and FeCl$_3$ from the film surface.

\subsection*{Conductivity Measurements}
Conductivity was measured in van der Pauw configuration,\cite{Chwang:1974,Koon:1989} following the method used in our previous work.\cite{Jacobs:2021a}. All measurements were performed in a nitrogen glovebox  Measurements were performed on a Karl Suss probe station inside a nitrogen glovebox (\textless 20ppm O$_2$) using an Agilent 4155B sourcemeter. Hysteresis I-V curves were measured with current sourced along each set of neighboring electrodes. This routine generates several redundant data points, enabling us to verify that hysteresis, current reversal, and reciprocity ($\frac{V_{12}}{V_{34}} = \frac{V_{34}}{V_{12}}$) remain below 3\%, in line with NIST recommendations.\cite{nist} Contact size effects contribute \textless1\% to the relative error,\cite{Chwang:1974,Koon:1989} therefore the uncertainty in conductivity is generally dominated by the thickness uncertainty (Bruker Dektak XT). As in our previous work, conductivity and carrier density are calculated from the undoped film thickness to ensure that the variation conductivity between samples is proportional to a change in the charge transport properties of the polymer chains.

\subsection*{UV-vis spectroscopy}
UV-vis-NIR spectra were measured using a Shimadzu UV-3600i dual beam spectrometer (3 nm monochromator width; 2 nm data interval), and background subtracted from separate measurement of uncoated substrates. Noise reduction was performed in the IR (\textless 0.75 eV) and UV (\textgreater 3.02 eV) regions (Savitzky-Golay filter) as previously reported\cite{Jacobs:2021a} to improve signal to noise when fitting the UV region.

\subsection*{XPS Measurements}
XPS spectra were collected on a Thermo Scientific Escalab 250xi. For the PBTTT samples  a pass energy of 20 eV, step size of 0.1 eV, spot size of 400 $\mu$m were used, and 30 scans were recorded for each sample. During the measurements of P3HT, DPP-BTz, and IDTBT instrument issues reduced the SNR required a larger spot size of 900 $\mu$m and 170 scans per spectrum. To minimize charging, films were prepared on gold electrodes and the flood gun was used. Data was processed using CasaXPS software.\cite{Fairley:2005} A Shirley background was used in all fits.\cite{Shirley:1972} Sulfur 2p spectra are characterized by a doublet (2p$_{3/2}$ and 2p$_{1/2}$) with 2:1 area ratio and spin-orbit coupling $\Delta$ = 1.18 eV.\cite{Moulder:1992} These constraints were enforced during all fits; linewidth was allowed to vary within reasonable ranges. A Voigt lineshape was assumed. Error bars on atomic concentrations were estimated by a Monte Carlo process in CasaXPS. 

\subsection*{QNMR Measurements}
Ion exchange doped PBTTT films (100 or 300 sec exposure time, 100/1 mM BMP TFSI / FeCl$_3$ in AN) were prepared on 2 cm square glass slides. The outer 1 mm of the film was removed to eliminate any thickness non-uniformities from spin coating, leaving a 1.8 $\pm$ 0.05 cm square film. This film was dedoped using a 10\% v/v triethylamine (TEA) CD$_3$CN solution. After 5 minutes of dedoping, the solution was removed by syringe and dispensed into an NMR tube; additional dedoping solution was used to remove any residue left on the film surface and in the syringe. A $^{19}$F QNMR standard, 4,4-difluorobenzophenone (DFBP) was then added to each tube (510 nmol, as 20 $\mu$L of a 25.49 mM standard solution) and mixed well. Spectra were acquired on a Bruker Avance III spectrometer (400 MHz, 9.4 T) with inverse gated $^1$H decoupling used for $^{19}$F spectra. 64 scans were acquired with a 45 s recycle delay, informed by a preceeding $^{19}$F T$_1$ measurement (TFSI 1.59 s; DFBP 4.35 s). The spectra were referenced to DFBP at -109 ppm. All peaks were integrated over a 15 Hz wide window centered on the peak position.

\subsection*{GIWAXS Characterization} 
Grazing-incidence Wide-angle X-ray Scattering (GIWAXS) characterization was done at the Advanced Photon Source (APS) Beamline 8-ID-E at Argonne National Laboratory. X-ray beam energy was 10.9 keV and incidence angle was 0.13$^{\circ}$. Two exposures of 2.5 second (5 s of exposure in total) were collected from each sample, recorded by a Pilatus 1 M detector located 228.16 mm from the sample. Data processing was performed using the MATLAB package GIXSGUI.\cite{gixsgui} Paracrystallinity and lattice parameters values were extracted by fitting linecuts to Gaussian functions plus an exponential background. The pi-stacking peak widths and positions was then used to calculate the $\pi-\pi$ paracrystallinity as:
\begin{equation}
    g = \frac{1}{2\pi}\sqrt{\Delta_q d_{hkl}}
\end{equation}
where $\Delta_q$ is the diffraction peak full width at half maximum, and $d_{hkl}$ is the interplanar distance. This expression assumes the pi-stacking coherence length is dominated only by paracrystalline disorder, which is generally understood to be the dominant type of disorder in conjugated polymers.\cite{Rivnay:2011a}

\subsection*{Conformational search procedure}

To investigate the supramolecular organization of PBTTT:TFSI systems, molecular mechanics (MM) and molecular dynamics (MD) calculations have been performed within the Materials Studio package.\cite{materialsstudio} A few years ago, we developed a Dreiding-based forcefield adapted to neat PBTTT.\cite{Venkateshvaran:2014} In this work, the same approach has been used and extended to TFSI. In particular, the atomic charges of TFSI have been set to the ESP charges calculated on a fully-optimized TFSI anion at the MP2/6-31G** level. Starting from the crystalline structure of PBTTT which contains one monomer unit, many different larger systems have been built by inserting TFSI anions between the alkyl chains or close to the PBTTT conjugated cores. In all cases, a PBTTT:TFSI ratio of 1:1 has been chosen as suggested by the experimental XPS/NMR characterisation at high doping levels. Given the anionic nature of TFSI, the atomic charges of the PBTTT conjugated cores have been rescaled to ensure electroneutrality; the positive excess charges are thus distributed evenly in the polymer chains, a reasonable approximation for the heavily-doped polaron lattices modeled here. 

The conformational search procedure to extract the most stable supramolecular organization involves four steps: (i) all starting structures are optimized at the MM level; (ii) 2ns-quenched MD runs (NPT, T = 300 K, quench frequency = 5 ps) are then performed on each optimized structure until the energy between two successive quenched systems no longer decreases; (iii) on the most stable structures obtained at step (ii), 2ns-quenched MD runs are performed at higher temperature, successively at 400K and 500K; (iv) quenched simulations (t = 2 ns), using as starting points the most stable structure of the last quenched systems in step (iii), are performed at increasing temperature (300K, 400K, and 500K) following the procedure developed in steps (ii) and (iii) to finally extract the most stable structure when the energy do not longer decrease between two successive cycles. 
The CASTEP module within the Materials Studio software has then been used to refine the most stable structure determined at the classical level. Geometry optimizations have been performed with the PBE functional and using the Grimme dispersion correction method, with all atomic positions and unit cell parameters allowed to vary.

\subsection*{GIWAXS pattern calculations}
When generating the GIWAXS patterns,\cite{Lemaur:2013} we have defined the x-y plane as the lamellar plane. The z direction therefore corresponds to the axis perpendicular to the lamellar plane. The angular position of the different spots are calculated by comparing the orientation of the different crystallographic planes as obtained from the Materials Studio Reflex module with respect to the x-y plane while the radial distance with respect to the origin characterizes the interplane distances.

However, in thin films, all crystallites do not have the same orientation with respect to the substrate and thus the spots are broadened depending on the amount of disorder present in the films. In our methodology, the intensity of a plane oriented with an $\Phi_n$ angle with respect to the x-y plane and corresponding to a peak at 2$\Theta_n$ is pondered by a Gaussian function whose standard deviation $\sigma$ can be varied in order to reproduce the different degrees of disorder in the film. The pondered intensity In is written as:
\begin{equation}
I_n=I_{n0} \frac{1}{\sigma \sqrt{2 \pi}} exp(\frac{-(\Phi_n^2)}{2\sigma^2 })
\end{equation}
An instrumental broadening of the peaks In was then introduced by a Lorentzian function independent of 2$\Theta$, in such a way that the intensity $I$ of the pattern at 2$\Theta$ is:
\begin{equation}
I=\sum_n \frac{1}{ \frac{((2\Theta-2\Theta_n )^2}{\Delta^2}} I_n
\end{equation}				
The broadening is adjusted by the parameter $\Delta$ to match the experimental peak width.

\subsection*{Ionic size calculations}
The TeraChem package~\cite{terachem1, terachem2}, version 1.9, was used to perform density functional theory calculations of the electronic structure for all anions. We used the B3LYP functional~\cite{B3, LYP} with the Grimme D3 dispersion correction~\cite{grimmeD3}, and the 6-311G++(d,p) basis set. Initial molecular structures were generated with the Avogadro package~\cite{avogadro} version 1.2.0, and were pre-optimized using a UFF forcefield~\cite{UFF} prior to full geometry optimization with TeraChem~\cite{DLFIND}. The size of each anion is encoded by a metric called the gyration tensor,
\begin{equation}
    \tilde{R}^2_{\alpha\beta} = \frac{1}{N} \sum_{i=1}^{N} r_{i,\alpha}r_{i,\beta}
\end{equation}
where $r_{i,\alpha}$ is the $\alpha$ Cartesian component of the position of atom $i$. The smallest eigenvalue of this tensor, $\lambda_x$, is used as a measure of the shortest approach distance to the ionic center of mass.

\subsection*{Electronic structure calculations}
The electronic structure of doped polymers is described with a model for interacting spinless-fermions on a 2D lattice. 
The Hamiltonian reads
\begin{eqnarray} \label{e:Hmet}
H&=& \sum_{\langle i,j\rangle} t_{ij} \,  \left( c_i^\dagger c_{j}^{\phantom{\dagger}} + h.c. \right ) 
+ \sum_i V_i^{(ion)}\, \hat n_i^{\phantom{\dagger}} 
\nonumber \\
&& + \frac{1}{2}\sum_{i, j} V_{ij} \,  \hat n^{\phantom{\dagger}}_i \hat n^{\phantom{\dagger}}_j 
\end{eqnarray}
where $c_i^\dagger$ ($c_{i}^{\phantom{\dagger}}$) creates (annihilates) a particle at site $i$, 
$\hat n^{\phantom{\dagger}}_i =  c_i^\dagger c_i^{\phantom{\dagger}}$, and $t_{ij}$ are charge-transfer integrals.
$V_{i}^\mathrm{(ion)}$ and $V_{ij}=(\varepsilon_r | \mathbf{r}_j- \mathbf{r}_i |)^{-1}$ are the ionic potential and the Coulomb interaction, both screened by a dielectric constant $\varepsilon_r=3.5$.
The model is solved in the Hartree-Fock approximation on systems of $48\times14$ sites, accounting for periodic boundary conditions.
The model effectively accounts for the effect of low and high frequency vibrations and it is parameterized with experimental data and atomistic calculations. The conductivity has been evaluated in the framework of the transient localization theory.\cite{Fratini16}
Full detail on the model and its parameterization are provided in Supporting Information Section 9.

\bibliographystyle{unsrt}
\bibliography{IEx_bib}

\end{document}